\documentclass[10pt,conference]{IEEEtran}

\usepackage{amsmath,amssymb,amsfonts}
\usepackage{algorithmic}
\usepackage{graphicx}
\usepackage{textcomp}
\usepackage{xcolor}
\def\BibTeX{{\rm B\kern-.05em{\sc i\kern-.025em b}\kern-.08em
    T\kern-.1667em\lower.7ex\hbox{E}\kern-.125emX}}
    
\PassOptionsToPackage{table,xcdraw}{xcolor}

\AtBeginDocument{%
  \providecommand\BibTeX{{%
    Bib\TeX}}}

\usepackage[english]{babel}
\usepackage[utf8]{inputenc}

\usepackage{flushend}
\usepackage{balance}
\usepackage{multirow}

\usepackage{color}
\usepackage{booktabs}
\usepackage{fancybox}
\usepackage{xcolor}
\usepackage{float}
\usepackage{array,graphicx}
\usepackage[flushleft]{threeparttable}
\usepackage[most]{tcolorbox}
\usepackage{xspace}
\usepackage{listings}
\usepackage{pifont}
\usepackage{tikz}
\usetikzlibrary{shapes.geometric}
\usepackage{blindtext}
\usepackage{colortbl}
\usepackage[table,xcdraw]{xcolor}

\usepackage{makecell}
\usepackage{stfloats}
\usepackage{url}

\usepackage{colortbl} 
\usepackage[table,xcdraw]{xcolor}
\usepackage{booktabs}

\usepackage{tcolorbox}

\usepackage{comment}

\definecolor{purplish}{HTML}{D8D0E3}
\definecolor{purplishlight}{HTML}{EBE7F1}
\definecolor{purplishdark}{HTML}{875afb}

\newtcolorbox[auto counter,number within=section]{rqbox}[2]{
    nameref=#1,
    title=\small{#1}, 
    enhanced,
    attach boxed title to top left={yshift=-6pt, xshift=8pt},
    boxed title style={size=small,boxsep=1pt},
    colframe=purplishdark,colback=white,colbacktitle=purplishdark,
    boxsep=2pt,left=2pt,right=2pt,top=6pt,bottom=2pt,middle=2pt
}

\newcommand{\rqtextone}{How does OSSDoorway support students in contributing to OSS projects?}

\newcommand{\rqone}[2][]{
    \begin{rqbox}{\textbf{Research Question}}{#2}
        \rqtextone
        #1
    \end{rqbox}
}

\begin{document}

\title{OSSDoorway: A Gamified Environment to Scaffold Student Contributions to Open Source Software}


\author{\IEEEauthorblockN{Italo Santos\IEEEauthorrefmark{1}, Katia Romero Felizardo\IEEEauthorrefmark{2}, Anita Sarma\IEEEauthorrefmark{3}, Igor Steinmacher\IEEEauthorrefmark{1} and Marco A. Gerosa\IEEEauthorrefmark{1}}
\IEEEauthorblockA{\IEEEauthorrefmark{1}Northern Arizona University, Flagstaff, AZ, USA\\}
\IEEEauthorblockA{\IEEEauthorrefmark{2}Federal Technological University of Paraná, PR, Brazil\\}
\IEEEauthorblockA{\IEEEauthorrefmark{3}Oregon State University, Corvallis, OR, USA\\}
Email: italo.santos@nau.edu, katiascannavino@utfpr.edu.br, anita.sarma@oregonstate.edu\\igor.steinmacher@nau.edu, marco.gerosa@nau.edu}

\maketitle

\begin{abstract}
    Software engineering courses enable practical learning through assignments requiring contributions to open source software (OSS), allowing students to experience real-world projects, collaborate with global communities, and develop skills and competencies required to succeed in the tech industry. 
     Learning software engineering through open source contribution integrates theory with hands-on practice, as students tackle real challenges in collaborative environments. 
    However, students often struggle to contribute to OSS projects and do not understand the contribution process. Research has demonstrated that strategically incorporating game elements can promote student learning and engagement.
    This paper proposes and evaluates OSSDoorway, a tool designed to guide students contributing to OSS projects.
    We recruited 29 students and administered a self-efficacy questionnaire before and after their use of OSSDoorway, along with qualitative feedback to assess challenges, interface features, and suggestions for improvement. 
    The results show that OSSDoorway boosts students' self-efficacy and provides a structured, gamified learning experience. Clear instructions, real-time feedback, and the quest-based system helped students navigate tasks like using GitHub features to submit pull requests and collaborating with the community. 
    Our findings suggest that providing students with a supportive gamified environment that uses feedback and structured quests can help them navigate the OSS contribution process.
\end{abstract}

\textbf{\textit{keywords: open source software, gamification, software engineering education.}}

\section{Introduction}
\label{sec:introduction}

Teaching software engineering (SE) is challenging~\cite{sedelmaier2012research, cantone2014agile}. Traditional SE courses tend to emphasize theoretical methodologies and concepts, offering a limited focus on preparing students to work with real-world, complex software systems~\cite{holmes2018dimensions, cawley2014incorporating}. To better align education with industry practices, SE courses should go beyond teaching concepts, methods, and techniques to include practical skills and attitudes that reflect the current software development landscape~\cite{nascimento2013using}. One approach that bridges this gap is integrating student participation in Open Source Software (OSS) projects as part of the SE curriculum~\cite{nascimento2013using, pinto2017training}. OSS projects foster collaborative environments where a community works together to develop software systems, providing a real-world context for learning.

Supporting students to contribute to OSS projects helps to prepare the future SE workforce~\cite{pinto2017training, silva2020google, pinto2019training}. By contributing to OSS, students at the beginning of their careers gain practical experience in both technical and soft skills, enhancing their confidence when pursuing industry positions~\cite{nascimento2013using, pinto2019training, braught2018multi, morgan2014lessons}. Successful contributions to OSS projects raise students' visibility among peers~\cite{cai2016reputation, riehle2015open} and help secure jobs \cite{Singer2013CSCW, Marlow2013CSCW}. Additionally, OSS contributions serve society beyond personal and project-level benefits by improving widely-used software~\cite{riehle2015open, parra2016making, greene2016cvexplorer}. 

However, contributing to OSS is not easy, and students usually feel discouraged by various barriers that hinder their participation~\cite{pinto2017training, steinmacher2015social, mendez2018open, padala2020gender, trinkenreich2021women, steinmacher2016overcoming, santos2022hits}, and need orientation. 
Gamification is a promising strategy for engaging students by incorporating game elements into educational systems to enhance motivation~\cite{dichev2017gamifying}. By leveraging gamification, students are more likely to stay committed, enjoy their contributions, and gain educational benefits~\cite{pedreira2015gamification, dicheva2015gamification, bartel2016gamifying}. For instance, Diniz et al.~\cite{PS04diniz2017using} implemented gamification strategies in GitLab to boost motivation and collaboration among undergraduate students contributing to OSS projects. Their findings indicate that game elements like quests and points help sustain engagement while guiding and motivating students to participate. In this paper, we designed and evaluated OSSDoorway, a tool to support students' contributions to OSS projects by leveraging a gamified environment to scaffold students' contributions. We define scaffolding following the description in~\cite{Belland2014} as the support provided by a teacher, parent, peer, or a computer- or paper-based tool, enabling students to engage in and develop proficiency in a task they would otherwise be unable to complete independently.
The following research question guided our research:

\rqone{}

To address this research question, we followed two steps.
In \textbf{Step 1, OSSDoorway Design and Development}, we conducted requirements elicitation, followed by the tools design, formative studies, and expert evaluations through an iterative process. Moving to \textbf{Step 2, Summative Evaluation}, we analyzed the data of 29 students who took the Open Source Software Development course and underwent a self-efficacy assessment before and after using OSSDoorway. We requested students' feedback, asking them to identify challenges they faced during quests, a key gamification element of OSSDoorway, point out the features they found most useful, and suggest improvements for the tool.

This paper makes three primary contributions: (1) OSSDoorway, a gamified tool designed to support students in contributing to OSS projects; (2) an empirical evaluation of OSSDoorway, demonstrating its impact on enhancing students' self-efficacy and ability to complete OSS-related quests; and (3) insights into students' perceptions of the proposed approach, highlighting the benefits and challenges they experienced and suggesting areas for improvement.

Our findings indicate that OSSDoorway supports students in contributing to OSS projects by enhancing their self-efficacy and offering a structured, gamified learning experience. OSSDoorway was particularly effective for women, who initially demonstrated lower levels of self-efficacy than their men counterparts. After using the tool, the difference in self-efficacy was no longer statistically significant. The integration of clear instructions, real-time feedback, and a quest-based system helped students successfully navigate complex tasks, such as using GitHub features, submitting pull requests, and collaborating with the community. These results indicate that OSSDoorway is a valuable tool for empowering students to contribute to OSS projects and evidence the effectiveness of gamification in this context.

\section{Background and related work}
\label{sec:backrelated}

This section outlines the background and related work involving gamification and its use in SE education.


\textbf{Gamification.} Dichev and Dicheva~\cite{dichev2017gamifying} describe gamification as a strategy to enhance student motivation in educational settings by integrating game elements. Gamification is increasingly being utilized in education to enhance the learning process~\cite{deterding2011gamification, sailer2020gamification}. Its key benefits include simplifying complex topics~\cite{marin2018empirical, ayub2019gamification}. For example, Deterding et al.~\cite{deterding2011gamification} propose that gamification simplifies complex material, making learning more accessible and promoting deeper understanding. They argue that games can enhance educational experiences by making them more interactive and effective for students. 
Toda et al.~\cite{toda2019analysing} proposed a framework for gamification strategies in educational settings based on a literature review and an evaluation with gamification and education experts. The resulting taxonomy included the description of 21 game elements, which we used as a baseline for the design of OSSDoorway game elements.


\textbf{Gamification for SE education.} Gamification has been effectively applied to SE education, including agile process~\cite{prause2012field}, software testing~\cite{rojas2016teaching, garaccione2022gerry, ozturk2022gamification}, design pattern~\cite{bartel2016gamifying}, software project management~\cite{annunziata2024serge, navarro2007comprehensive}, and business processes~\cite{garaccione2024gamification}. These applications highlight the broad potential of gamification to enhance educational experiences in the field. Su~\cite{su2016effects} developed a gamified framework to assess the impact of gamification on teaching SE, finding that gamified methods increased student motivation and improved academic performance. Similarly, Sheth et al.~\cite{sheth2012increasing} showed that incorporating gamification into a SE course enhanced student engagement in areas such as documentation, bug reporting, and testing. In the context of OSS, gamification strategies have been used to encourage contributions, such as through the OpenRank network algorithm and a monthly contribution leaderboard~\cite{zhao152023motivating}. Santos et al.~\cite{santoslr} identified software solutions that facilitate the onboarding of newcomers in software projects. Among these solutions, some used gamification techniques with newcomers, fostering engagement and boosting motivation~\cite{PS04diniz2017using, PS17toscani2018gamification}. Diniz et al.~\cite{PS04diniz2017using} implemented four game elements, i.e., quests, points, ranking, and levels in GitLab, and assessed their ability to motivate students to overcome orientation barriers. Although their approach shares similarities with this study, our work differs by conducting a systematic user-centered design of a more comprehensive gamified environment to support students' contributions in GitHub, employing a bot that interacts with students and updates the environment.  


\section{Method}
\label{sec:methodology}

\begin{figure*}[!bth]
    \centering
    \includegraphics[width=0.6\textwidth]{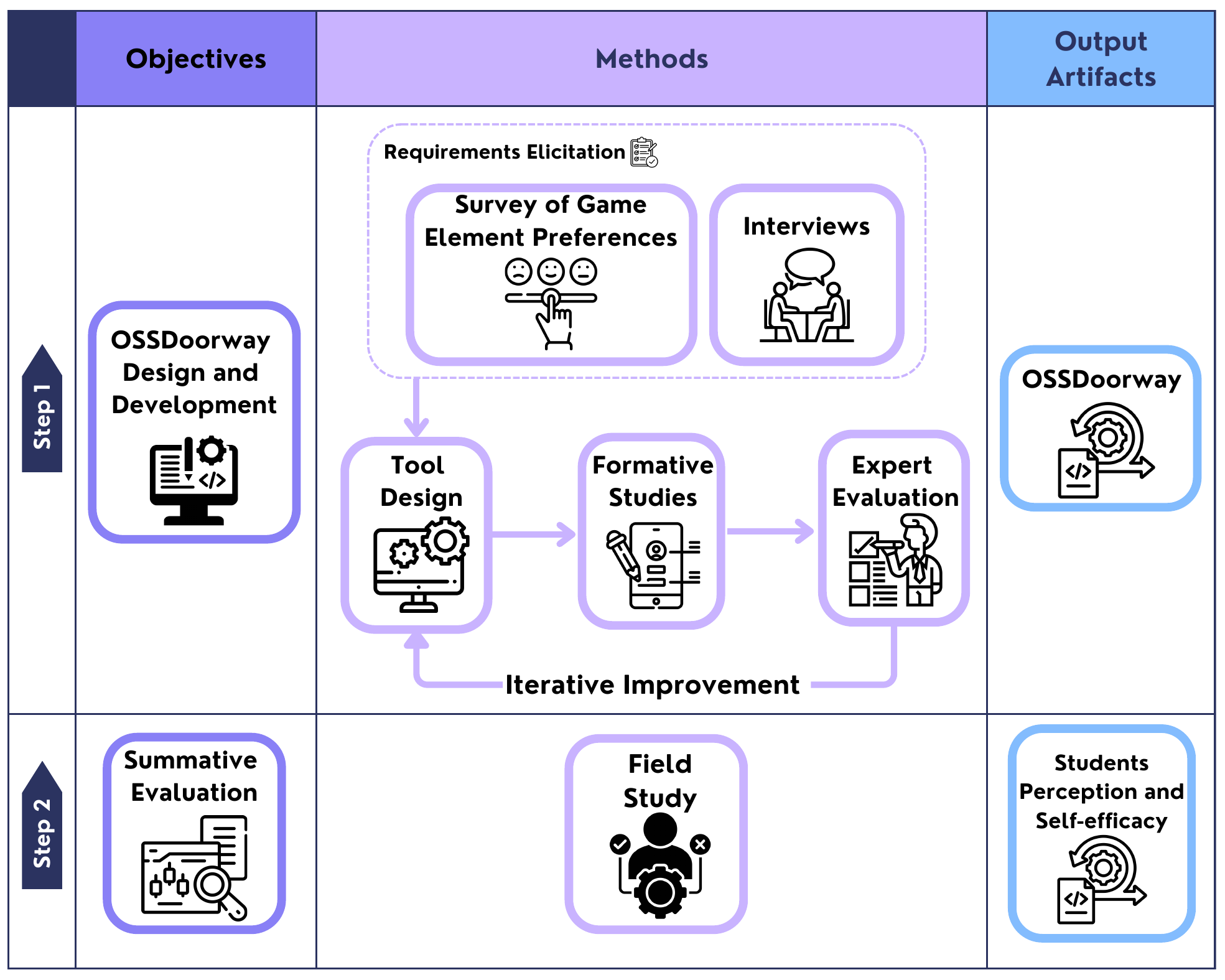}
    \caption{Method overview.}
    \label{fig:methodoverview}
\end{figure*}

This section outlines our approach to developing and evaluating the OSSDoorway tool. We divided the process into two steps: (i) OSSDoorway design and development and (ii) Summative evaluation, as observed in Figure~\ref{fig:methodoverview}.

\textbf{Step 1 -- OSSDoorway design and development.} First, we carried out the requirements elicitation through a survey to gather student perceptions of game elements to be implemented in OSSDoorway (published elsewhere~\cite{santosgamelements}). Then, we conducted interviews to further validate our findings and select the game elements to be implemented. Afterward, we followed an iterative user-centric design process that involved designing the tool, conducting formative evaluations, and assessing the proposed tasks with experts, i.e., instructors with experience teaching about OSS. In this process, we created and refined paper prototypes, which evolved into high-fidelity prototypes. Subsequently, we developed the actual tool. More details in Section~\ref{sec:ossdoorway}.

\textbf{Step 2 -- Summative evaluation.} We conducted an evaluation with 29 undergraduate students to assess OSSDoorway. In the study, students completed three different quests, each consisting of 3 to 5 different tasks. We administered a self-efficacy questionnaire before and after the students used OSSDoorway. The pre- and post-questionnaire responses were compared using the Wilcoxon signed-rank test~\cite{rosner2006wilcoxon}. At the end of the study, participants were asked to describe any difficulties they encountered while completing tasks, what aspects of the interface were most helpful, and provide suggestions for improving OSSDoorway. We qualitatively analyzed participants' responses to the open-ended questions using open coding procedures~\cite{strauss1998basics}. Section~\ref{sec:labstudy} provides additional details about this process.

\section{OSSDoorway design and development}
\label{sec:ossdoorway}

This section describes the OSSDoorway design and development process, including the requirements elicitation, tool design, formative studies, and expert evaluation.


\subsection{Requirements elicitation}
\label{sec:requirements}

In previous work~\cite{santosgamelements}, we surveyed 115 computer science students to identify their preferences for various game elements extracted from the taxonomy proposed by Toda et al.~\cite{toda2019analysing}. The survey was designed to provide insights from a user-centered design perspective before implementing the tool. In the survey, we tailored the description of the game elements within the context of the OSS contribution process.

The findings revealed that performance-oriented game elements, such as \textit{quests, stats, maps, levels, points, progress bars,} and \textit{badges}, were favored according to students. These elements offer valuable feedback on performance and foster a sense of progression and achievement. We paid particular attention to how perceptions of game elements varied across demographic groups, and we found no significant difference. Further details on this study can be found in~\cite{santosgamelements}.

After this preliminary work, we recruited eight computer science students for observation sessions. Participants in the observation sessions are designated as OB$<$X$>$, where X represents their assigned participant number. The participants comprised four undergraduate and four graduate students (3 women and 5 men). With their consent, the sessions were audio-recorded for note-taking purposes. During the sessions, we observed how the participants interacted with GitHub while performing two key tasks: (i) editing the Readme file and submitting a pull request and (ii) selecting and solving a non-code contribution task. These tasks were chosen because they exercise part of the process to contribute to an OSS project~\cite{steinmacher2016overcoming}.

We employed the think-aloud method~\cite{eccles2017think}, where participants openly verbalized their decisions and reasoning. This allowed us to ask follow-up questions during the observation to clarify their decision-making process, providing valuable insight into how GitHub facilitates student contributions to OSS projects. Each session lasted approximately one hour.

During the interviews, we asked participants about the possibility of using game elements to support and contribute to a project on GitHub. Overall, the interviewees recommended including the same game elements identified in the survey while highlighting additional important elements, such as streaks. For instance, OB1 mentioned, ``\textit{I like quests a lot because they are fun, and I expect the system to have streaks to keep me engaged}'', and OB4 noted, ``\textit{I would like a streak feature in the environment to give more points to players who access the system every day}''. OB3 also emphasized, ``\textit{I like how a gamified environment breaks complex tasks into simpler steps through quests}''. 

\begin{tcolorbox}[colback=white!50, colframe=cyan, title=\textbf{Summary of Requirements Elicitation}, fonttitle=\bfseries]

After the survey and the interviews, we identified the following features that OSSDoorway needs to provide:

\textbf{OSSDoorway features:}

\begin{itemize}
    \item Implement game elements that provide performance-oriented feedback: \textit{stats, maps, levels, points, progress bars,} and \textit{badges}.
    \item Implement streaks that allow users to earn additional rewards.
    \item Implement quests to support students in accomplishing the contribution goals.
\end{itemize}

\end{tcolorbox}

\subsection{Tool design}


OSSDoorway was designed to guide students through a series of quests, each aligned with specific OSS contribution goals to ensure students gain practical knowledge and skills in contributing to OSS projects. We established three contribution goals based on previous research on the barriers that newcomers face when contributing to OSS projects, with emphasis on the key stages of the contribution process~\cite{steinmacher2016overcoming}. Through these quests, students should be able to (i) understand the GitHub contribution process, (ii) interact with other project members, and (iii) make a contribution. 

To address goal \textit{(i) understand the GitHub contribution process}, we designed \textbf{Quest 1: Exploring the GitHub World}. This quest familiarizes students with GitHub essential features by guiding them through the following tasks:

\begin{itemize}\small
    \item Task 1 -- Explore the issue tracker;
    \item Task 2 -- Explore the pull-request;
    \item Task 3 -- Explore the fork;
    \item Task 4 -- Explore the README file;
    \item Task 5 -- Explore the project contributors;
\end{itemize}


To address \textit{goal (ii) interact with other project members}, we designed \textbf{Quest 2: Introducing Yourself to the Community}, which focus on guiding students on how to communicate with project contributors, as described below:

\begin{itemize}\small
    \item Task 1 -- Choose an issue to work on;
    \item Task 2 -- Assign your user[id] to work on the issue;
    \item Task 3 -- Post a comment in the issue introducing yourself;
    \item Task 4 -- Mention a contributor to help you to solve the issue;
\end{itemize}


Finally, for the third contribution goal, \textit{(iii) make a contribution}, we designed \textbf{Quest 3: Making Your First Contribution}. This quest involves students completing a non-code contribution, as described below:

\begin{itemize}\small
    \item Task 1 -- Solve the Issue (non-code contribution) and Submit a Pull Request;
    \item Task 2 -- Post in the issue asking for someone to review it;
    \item Task 3 -- Close the issue;
\end{itemize}

Through this process, students gain hands-on experience with the contribution workflow in GitHub. For the first version of OSSDoorway, we concentrated on non-code contributions to accommodate users without a technical background and to focus on the process and environment challenges, which are common across projects. The literature has shown that non-code contributions are a great way to join OSS~\cite{trinkenreich2020hidden}. We plan to expand OSSDoorway to include code review and refactoring quests.

\begin{figure}[!ht]
    \centering
    \includegraphics[width=8cm]{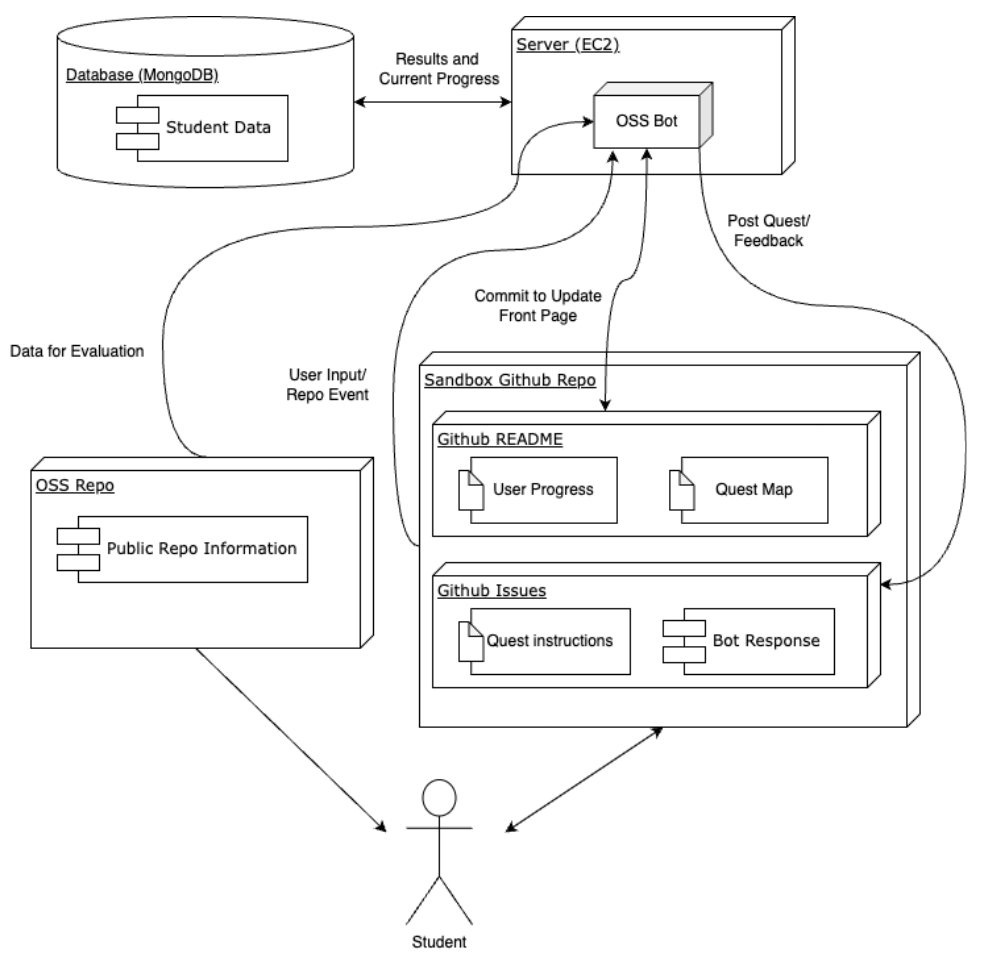}
    \caption{OSSDoorway Design Overview.}
    \label{fig:diagram}
\end{figure}

\textbf{Tool description.} The students interact with the sandbox GitHub repository, where they can monitor their progress through the GitHub README, which serves as the ``home'' page, displaying all available tasks and user progress. Students develop and test their contributions in isolated sandbox environments, providing a safe learning space to experiment, make mistakes, and refine their code without impacting the original OSS projects. Task completion is guided by quest instructions provided in GitHub Issues, where the system facilitates interaction and provides feedback through a bot implemented to check students' responses (OSS Bot). The OSS Bot verifies the student's actions, updates the sandbox repository, and records progress in a MongoDB database. These updates trigger changes that are continuously reflected on the front page while generating feedback for students and collecting evaluation data to assess the tool's effectiveness. In addition to working within the sandbox environment, the student can interact with a public OSS repository, contributing to real-world OSS projects and gaining hands-on experience. The OSS bot communicates with the user through two GitHub pages: the main repository page, where the README is visible, and GitHub Issues, where the quest interactions occur. OSSDoorway has been developed with a NodeJS~\cite{nodejs} backend that utilizes the Probot~\cite{probot} framework to create a bot for GitHub, making the solution more approachable for managing and automating GitHub tasks. 
Figure~\ref{fig:diagram} illustrates the key components of the OSSDoorway tool and their interactions. 

\begin{figure*}[!ht]
    \centering
    \includegraphics[width=1.0\textwidth]{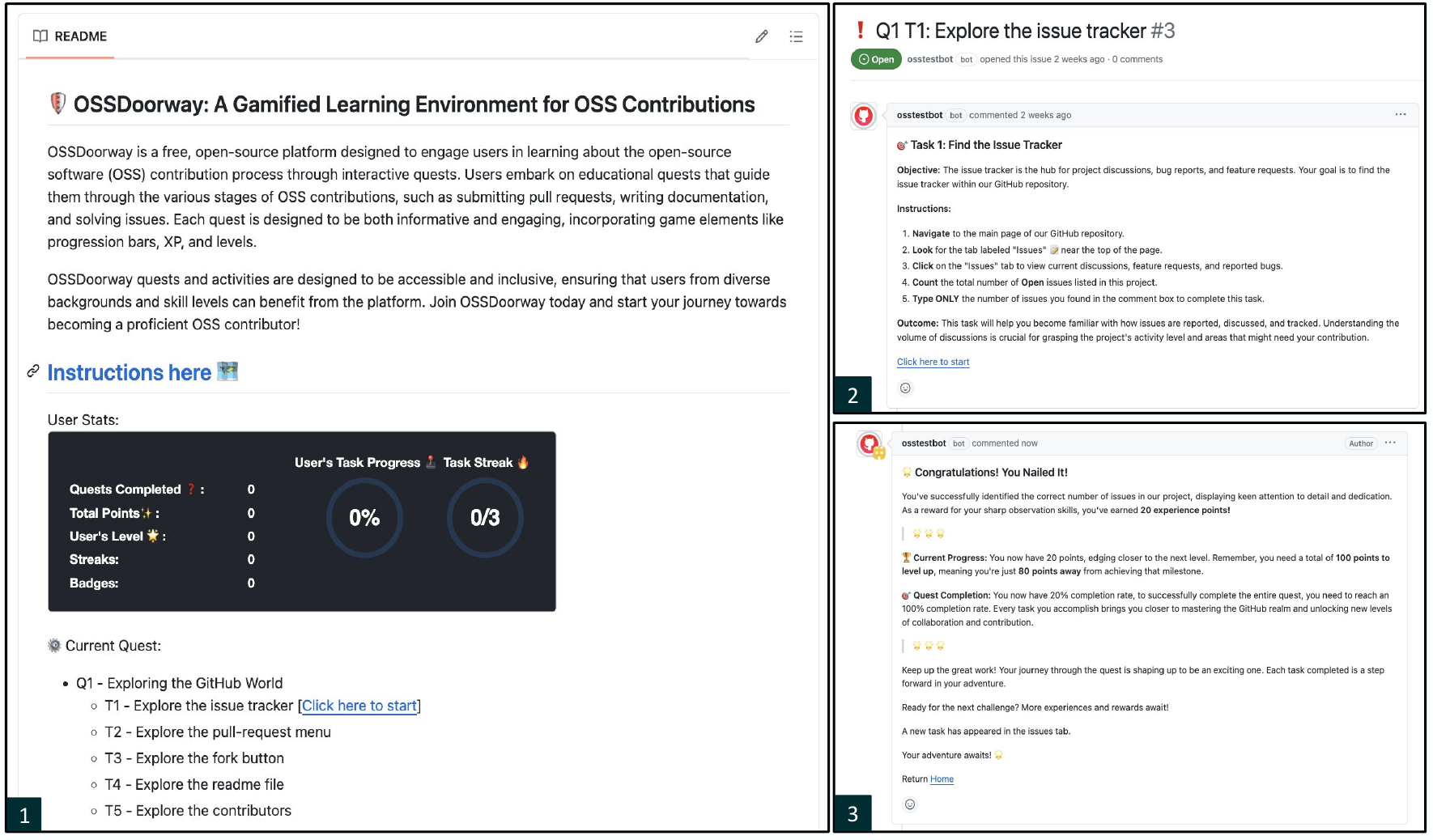}
    \caption{The left side (1) displays the OSSDoorway README page, where students can track their progress and view available tasks along with game elements such as progress bars, experience points, levels, badges, and streaks. The top right side (2) shows the GitHub Issues page with detailed instructions for Task 1. The bottom right side (3) presents feedback from the OSS Bot after the student successfully completes the task, providing progress updates and awarding experience points.}
    \label{fig:ossdoorwaypages}
\end{figure*}

The interface of the OSSDoorway tool enables students to track their progress, view available quests and tasks, and interact with game elements such as progress bars, experience points, levels, badges, and streaks, as shown in Figure~\ref{fig:ossdoorwaypages}. In this figure, the task description for Quest 1 is displayed, where students are instructed to explore the issue tracker. Finally, the bot's feedback is presented, and the student points are awarded upon successful task completion.

\textbf{OSSDoorway gamification elements.} OSSDoorway implements eight game elements previously identified in the requirement elicitation (Section \ref{sec:requirements}). Students can track their \textit{Quests completed} while earning \textit{Experience points} based on task difficulty. As they accumulate points, they \textit{Level up}, unlocking new quests and rewards. \textit{Badges} are awarded for completing quests and maintaining \textit{Streaks}, earned by completing three tasks in a row. The tool also provides a snapshot of \textit{Student stats}, summarizing completed quests, total points, and current level. At the same time, a \textit{Map} helps students visualize their journey, and a \textit{Progress bar} shows students' progress. Moreover, OSSDoorway includes a correctness check through the integrated OSS Bot, which assesses whether students have completed a task and are ready to proceed to the next one.

To make OSSDoorway accessible for users with different cognitive styles, we employed directives suggested by Burnett et al.~\cite{burnett2018gendermag}. For example, we emphasized the practical impact of completing tasks, appealing to purpose-driven users while providing enjoyment for those \textit{motivated} by technology. We considered comprehensive and selective \textit{information processing styles}, offering detailed instructions for those who prefer step-by-step guidance while allowing others to explore the GitHub environment more flexibly. To support users with lower \textit{computer self-efficacy}, OSSDoorway provides encouraging feedback and achievable tasks. It also addresses \textit{risk aversion} by allowing students to experiment with low-stakes tasks, guided by the OSS Bot, which ensures mistakes can be corrected. Finally, OSSDoorway supports structured, process-oriented \textit{learning style} as well as tinkering, offering a flexible environment with clear instructions.

\subsection{Formative studies}

\textbf{OSSDoorway prototypes.} We created paper prototypes of OSSDoorway and conducted two evaluation rounds. Participants in the prototype sessions are designated as PS$<$X$>$, where X represents their assigned participant number. With their consent, the sessions were audio-recorded for note-taking purposes.

During the first round, the group consisted of one undergraduate student, two graduate students, and one industry professional, all men. Overall, participants highlighted the need for clearer instructions to navigate the environment. PS1 emphasized that clear explanations for each game element were essential for understanding OSSDoorway. PS3 noted that some quests were unclear and required rewording or more detailed descriptions to clarify user actions. PS2 found the navigation confusing and suggested adding visual cues for better guidance. Based on this feedback, we revised the prototype by adding instructional guidelines on interacting with OSSDoorway and improving task descriptions to provide feedback.

During the second round of tests, the group was composed of four graduate students (one woman and three men). Participants indicated more satisfaction with the prototype. They suggested that they would like to see some competition elements in the environment. Moreover, PS5, PS7, and PS8 indicated that competitive aspects, such as leaderboards or rewards, could significantly enhance motivation. Specifically, PS3 and PS7 desired a more competitive system, suggesting rewards like GitCoins or leaderboards similar to those found in Duolingo, a popular gamified language-learning app. Furthermore, P08 also suggested real-world rewards as an incentive.

After refining these two paper prototypes, we developed a high-fidelity prototype, as shown in Figure~\ref{fig:proto3}. The paper prototypes are available in the supplemental material~\cite{replicationpackage}. In the high-fidelity prototype, we incorporated all participant suggestions, except for the competitive elements, which were excluded based on survey results indicating that competition and pressure were viewed less favorably~\cite{santosgamelements}.

\begin{figure}[!ht]
    \centering
    \includegraphics[width=8cm]{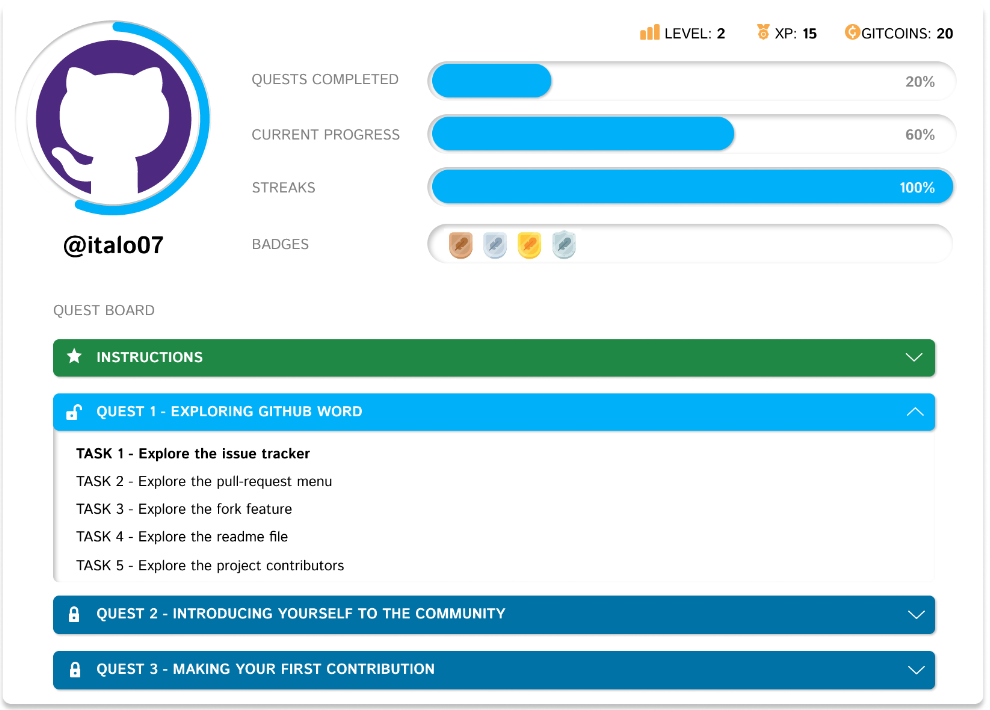}
    \caption{OSSDoorway Page - High Fidelity Prototype.}
    \label{fig:proto3}
\end{figure}

\subsection{Expert evaluation}


During the development of OSSDoorway, we evaluated the quests included in OSSDoorway to ensure that they align with our intended three contribution goals defined previously in Subsec.~\ref{sec:requirements}. To assess how well the proposed quests matched the contribution goals, we interviewed twelve experts with experience teaching OSS in their computer science curriculum. Among the 12 participants, all hold doctoral degrees and have significant experience in SE and OSS development, with some specializing in Human-Computer Interaction (HCI) and Artificial Intelligence (AI). Over half of the participants have more than 20 years of teaching experience. Their experience with OSS started as early as 2001. Seven out of twelve participants are full professors at well-known universities, teaching courses related to OSS development, software engineering, and capstone projects. 

Overall, the experts' feedback was highly positive. Several participants noted that the quests closely mirrored the tasks they already assign to students in their courses, reinforcing that the \textit{contribution goals are aligned with education goals} in the classroom and that the proposed quests support students in contributing to OSS projects. Experts also offered suggestions for improvement, such as reordering the tasks in quest 1 to place the exploration of the README file first, and including quests that require technical knowledge. We plan to incorporate this feedback in a second development cycle, adding more advanced quests to OSSDoorway.


\section{Summative evaluation}
\label{sec:labstudy}


We adopted OSSDoorway over two editions of an OSS development course at a US university, each led by a different instructor and attended by master's students. Our study evaluated how OSSDoorway supports students in contributing to OSS projects. We administered a self-efficacy questionnaire before and after the students used the platform, using a 5-point Likert scale for each question. Additionally, we gathered qualitative feedback at the end of the study by asking participants to describe any challenges they faced, the most helpful aspects of the interface, and their suggestions for improvement. 

\textbf{Pilot.} We conducted a pilot study with two researchers to gather feedback on the field study instruments (questionnaires) and overall design. We revised the questionnaires based on the feedback received, including removing certain questions to reduce the length of the study.

\textbf{Participants.} We applied OSSDoorway in the classroom, in two different semesters. Out of the 37 students who took the courses and used the environment, 29 agreed to have the data analyzed. The students ranged from 20 to 25 years old; 13 were women, and 16 were men. All students were enrolled in a Master's degree program in Information Technology or Computer Science. 
We asked participants about their experience with GitHub and OSS. While some mentioned having used GitHub before, upon further inquiry, it became clear that their interaction was limited to account creation without actual project contributions. Therefore, none had prior experience contributing to OSS and using GitHub. 

\textbf{Data analysis.} \textit{1) Likert scale questions:} We asked the level of agreement of participants using Likert scale questions that run from \textit{strongly disagree} (encoded as 1) to \textit{strongly agree} (encoded as 5), to rate their self-perception about their ability to contribute to OSS projects. The questionnaire was based on the work of Bandura~\cite{bandura2014social} and had seven questions. Participants answered those questions before and after the experiment. The goal was to capture the students' self-perceived efficacy. The questions were prefixed with ``I am confident that I can:'' followed by: 

\begin{itemize}
    \item [(i)] \ldots use GitHub issue tracker to find open issues; 
    
    \item [(ii)] \ldots understand GitHub pull-requests; 
    
    \item [(iii)] \ldots fork GitHub repositories; 
    
    \item [(iv)] \ldots find someone to help me with an issue using the GitHub web interface; 
    
    \item [(v)] \ldots open a pull request using GitHub web interface; 
    
    \item [(vi)] \ldots find an issue to work on and assign it to me;  
    
    \item [(vii)]\ldots use GitHub to contribute to projects.
\end{itemize}

\textit{2) Open questions:} We used open questions to ask participants to describe difficulties they encountered while completing tasks, what aspects of the interface were most helpful, and to provide suggestions for improving OSSDoorway. We qualitatively analyzed participants' comments following open coding procedures~\cite{strauss1998basics}. The process was conducted using continuous comparison and discussion until reaching a consensus. Two researchers jointly analyzed the sets of answers to establish common ground, discussing the applied codes and disagreements until reaching a consensus. Finally, a third researcher inspected the classification.

\subsection{Results}

All participants successfully completed all quests within the study time, with no discernible differences among demographics. Therefore, we answer RQ: \textit{\rqtextone} by analyzing (1) the results of the self-efficacy questionnaire, which was administered before and after the student's interaction with OSSDoorway, and (2) the students' perceptions of using OSSDoorway.

\subsubsection{Students' self-efficacy}

\begin{figure*}[!ht]
    \centering
    \includegraphics[width=1.0\textwidth]{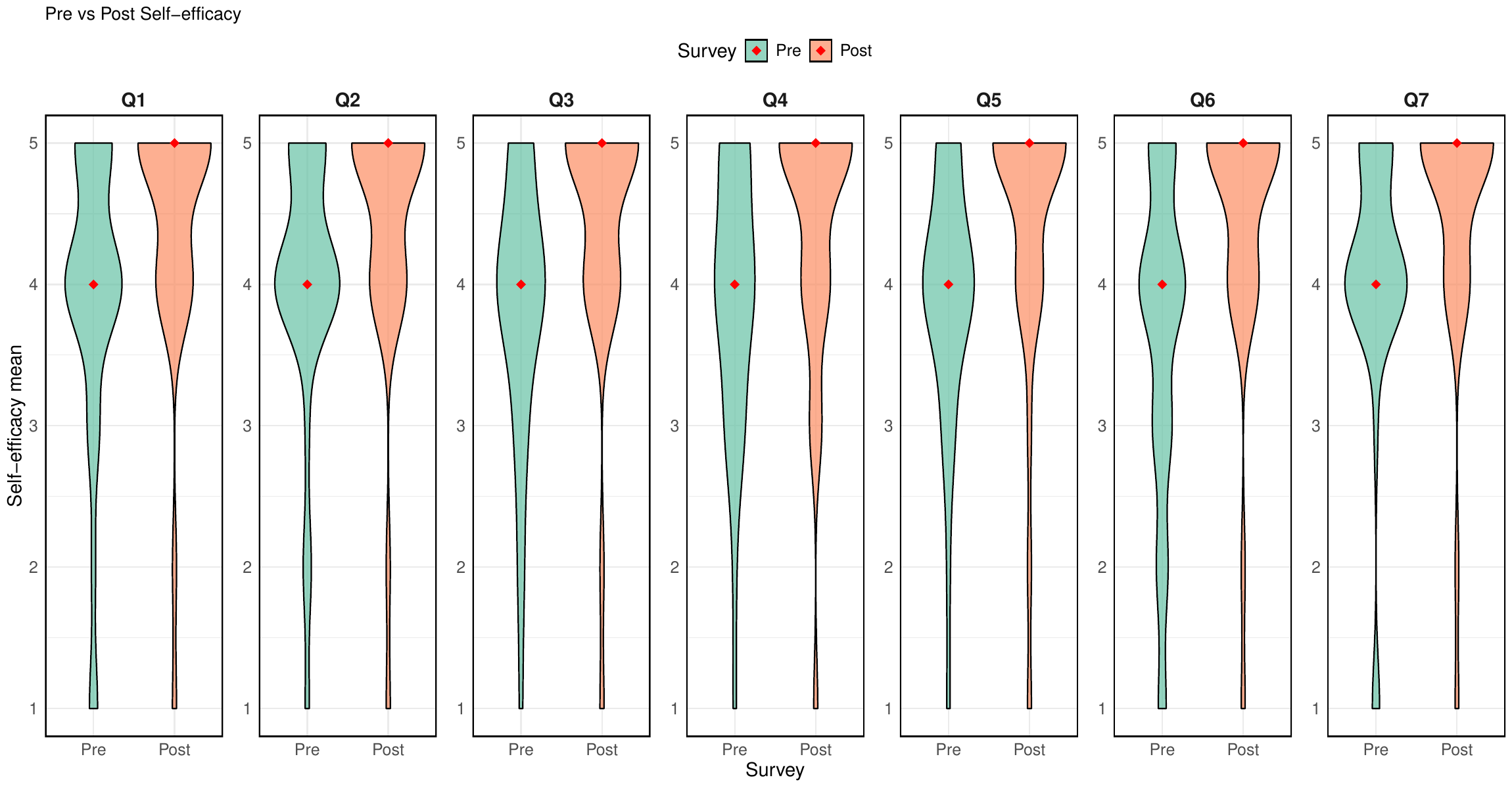}
    \caption{Responses to the 5-point Likert scale questions for the self-efficacy questions. The left side (``Pre'') shows students' responses to the questions before using OSSDoorway, while the right side (``Post'') shows their responses after using OSSDoorway. The red dot indicates the median response for each question. \\ \textit{\textbf{Question description:} I am confident that I can... \textbf{Q1:} use GitHub issue tracker to find open issues; \textbf{Q2:} understand GitHub pull requests; \textbf{Q3:} fork GitHub repositories; \textbf{Q4:} find help with an issue using the GitHub web interface; \textbf{Q5:} open a pull request using the GitHub web interface; \textbf{Q6:} find an issue to work on and assign it to myself; \textbf{Q7:} use GitHub to contribute to projects.} The left side (Pre) shows students' responses to the questions before using OSSDoorway, while the right side (Post) shows their responses after using OSSDoorway.}
    \label{fig:selfefficacy}
\end{figure*}

Figure~\ref{fig:selfefficacy} presents the results of the self-efficacy questionnaire that students filled out before (``pre'') and after (``post'') using OSSDoorway to contribute to OSS projects. Our findings show increased students' self-efficacy after using OSSDoorway, with post-scores consistently higher than pre-scores across all seven survey questions (Q1 to Q7). This indicates that students felt more confident in contributing to OSS projects after using OSSDoorway to complete the tasks. While the degree of improvement varied, the data suggests that OSSDoorway successfully enhanced students' self-efficacy in areas where they initially were less confident.

We calculated the Wilcoxon signed-rank test, illustrated in Table~\ref{tab:selefficacytable}, a frequently used nonparametric test for paired data (e.g., pre-and post-treatment measurements)~\cite{rosner2006wilcoxon}, and applied the Bonferroni correction for multiple comparisons. The difference in improvement (``pre'' vs. ``post'') using OSSDoorway is significant for four out of the seven tasks, which saw an increase in both mean and median scores, with all p-values below 0.05. The most substantial improvements were observed in tasks like finding help with issues (Q4) and assigning issues (Q6), which had the lowest pre-assessment scores and the most significant changes (p-values 0.024 and 0.003, respectively). These results suggest that OSSDoorway boosts students' confidence in navigating GitHub and contributing to OSS projects.

\begin{table}[htb]\scriptsize
\centering
\caption{Statistically significant differences (\(p < 0.05\)) in self-efficacy between pre- and post-assessment.}
\label{tab:selefficacytable}
\vspace{-4px}
\begin{tabular}{lrrrrr}
\Xhline{1pt}

 & \multicolumn{4}{c}{OSSDoorway} & \multicolumn{1}{l}{} \\ \cline{2-5}
 & \multicolumn{2}{c}{\begin{tabular}[c]{@{}c@{}}Pre \\ Self-efficacy\end{tabular}} & \multicolumn{2}{c}{\begin{tabular}[c]{@{}c@{}}Post\\ Self-efficacy\end{tabular}} & \multicolumn{1}{l}{} \\ \cline{2-6} 
 & \multicolumn{1}{l|}{mean} & \multicolumn{1}{l|}{median} & \multicolumn{1}{l|}{mean} & \multicolumn{1}{l|}{median} & \multicolumn{1}{l}{\(p\)-value$^{*}$} \\ \hline \hline
 
\multicolumn{1}{l|}{\begin{tabular}[c]{@{}l@{}}Q1...use GitHub issue \\ tracker to find open issues\end{tabular}} & \multicolumn{1}{r|}{3.9} & \multicolumn{1}{r|}{4} & \multicolumn{1}{r|}{4.4} & 5 & \multicolumn{1}{|r}{0.085} \\ \hline

\multicolumn{1}{l|}{\begin{tabular}[c]{@{}l@{}}Q2...understand \\ GitHub pull-requests\end{tabular}} & \multicolumn{1}{r|}{4} & \multicolumn{1}{r|}{4} & \multicolumn{1}{r|}{4.4} & 5 & \multicolumn{1}{|r}{\cellcolor[HTML]{98FB98}0.036} \\ \hline

\multicolumn{1}{l|}{\begin{tabular}[c]{@{}l@{}}Q3…fork GitHub\\ repositories\end{tabular}} & \multicolumn{1}{r|}{3.8} & \multicolumn{1}{r|}{4} & \multicolumn{1}{r|}{4.4} & 5 & \multicolumn{1}{|r}{\cellcolor[HTML]{98FB98}0.045} \\ \hline

\multicolumn{1}{l|}{\begin{tabular}[c]{@{}l@{}}Q4…find someone to help \\ me with an issue using the \\ GitHub web interface\end{tabular}} & \multicolumn{1}{r|}{3.9} & \multicolumn{1}{r|}{4} & \multicolumn{1}{r|}{4.4} & 5 & \multicolumn{1}{|r}{\cellcolor[HTML]{98FB98}0.024} \\ \hline

\multicolumn{1}{l|}{\begin{tabular}[c]{@{}l@{}}Q5...open a pull request \\ using GitHub web interface\end{tabular}} & \multicolumn{1}{r|}{3.9} & \multicolumn{1}{r|}{4} & \multicolumn{1}{r|}{4.4} & 5 & \multicolumn{1}{|r}{0.073} \\ \hline

\multicolumn{1}{l|}{\begin{tabular}[c]{@{}l@{}}Q6...find an issue to work \\ on and assign it to me\end{tabular}} & \multicolumn{1}{r|}{3.6} & \multicolumn{1}{r|}{4} & \multicolumn{1}{r|}{4.4} & 5 & \multicolumn{1}{|r}{\cellcolor[HTML]{98FB98}0.003} \\ \hline

\multicolumn{1}{l|}{\begin{tabular}[c]{@{}l@{}}Q7...use GitHub to \\ contribute to projects\end{tabular}} & \multicolumn{1}{r|}{4} & \multicolumn{1}{r|}{4} & \multicolumn{1}{r|}{4.5} & 5 & \multicolumn{1}{|r}{0.080} \\ \hline \hline

\multicolumn{6}{l}{$^{*}$\textit{\(p\)-value after applying the Bonferroni correction.}} \\

\Xhline{1pt}
\end{tabular}
\vspace{-10px}
\end{table}

\textbf{Segmented analysis.} We analyzed the data by segmenting it to compare men and women and applied the Mann-Whitney U test, a widely used nonparametric test for independent samples~\cite{macfarland2016mann}. We compared them in both the pre-and post-self-efficacy assessments (i.e., comparing men pre vs. women pre and men post vs. women post). The results revealed a statistically significant difference in the pre-test (p-value = 0.003), indicating that men had a higher self-efficacy before using OSSDoorway (mean = 4.09) compared to women (3.69). This result is consistent with the literature~\cite{burnett2018gendermag}, which found that women tend to have lower self-efficacy than men in their peer group. In the post-test, there was no significant difference between men and women (p-value = 0.914), with both groups achieving a median score of 5 and similar means (men: 4.5, women: 4.3). These findings suggest that while men initially exhibited higher self-efficacy, \textit{the use of OSSDoorway contributed to leveling self-efficacy for men and women}.

\subsubsection{Students' perception about OSSDoorway}

As mentioned, we adopted open coding in our qualitative analysis to identify recurring themes across students' responses.

\textbf{Difficulties using OSSDoorway.} We asked participants to share the most difficult aspects they encountered while using OSSDoorway. Feedback revealed that many students faced \textit{challenges to solve an issue} during Quest 3, which requires the student to solve an issue and submit a pull request. For example, P1 noted that ``\textit{solving the issue and submitting a pull request was a bit hard}'', while P9 shared that ``\textit{The most difficult is the implementation of pull request and committing the code to the main as it is my first time to learn and implement such thing.}'' Some participants also mentioned difficulties with navigation and task clarity. Overall, while some students found the process manageable with clear instructions—P25 remarked that ``\textit{The process was really easy. I didn't find anything difficult. The instructions were very detailed which helped me complete the quest on time.}''—others encountered specific challenges, particularly related to Quest 3 and navigating GitHub features.

\textbf{Tool support.} Our results highlight what students found helpful while using the OSSDoorway environment that supported them through task completion. One indicated feature was the README file in the OSSDoorway that shows to the user the game elements to help them track their progress, which participants frequently mentioned as crucial for their success. For example, P1 stated that ``\textit{accessing information through the README file was very helpful}''. The quest system and its process-oriented design also played a significant role in facilitating the learning experience. P25 noted that the quests helped them ``\textit{All the elements were easy to locate. The quest was interactive and was well defined}'', making the process easier to navigate. Similarly, P10 highlighted that the quests' structure provided a clear and logical progression, helping them track their progress effectively. Highlighted instructions within the tasks were also indicated as useful, P26 elaborated: ``\textit{I think the step-by-step instructions were very helpful}'', ensuring clarity and reducing confusion during task execution. Students also indicated that the OSS Bot offered real-time feedback and support. As P9 mentioned, ``\textit{Using of Chatbot to see if we have done the changes or not is the most helped part in the whole process for me.}''. At the same time, P11 emphasized ``\textit{Chat bot helped me a lot to confirm whether my answer is correct or not and description of each question of quest and directions in the question helped me to complete the assignment}''.

\textbf{Feedback to improve OSSDoorway.} The theme \textit{improve quests description} emerged in students feedback. P13 highlighted this with the remark: ``\textit{Even though the instructions were clear, it was a little difficult for a newbie to explore everything on his own.}'' This suggests that, while the instructions were adequate, a more detailed guide may be needed to better support students' needs. Overall, we received positive feedback, such as P2's comment: ``\textit{I think the process is as simple as it can be and I can't think of anything to specifically change to improve the process}'', and P25's observation: ``\textit{The quest can increase their level and add more interesting topics. It was really fun to learn about GitHub through this quest. I personally liked it.}'' These responses indicate that students found the experience using OSSDoorway enjoyable.

\rqone[
    \tcblower
    \textbf{Answer:} All students accomplished the tasks, and OSSDoorway significantly improved students' self-efficacy in contributing to OSS projects. Even though men had a higher self-efficacy than women before using OSSDoorway, this difference disappeared afterward. According to the students, OSSDoorway provides clear instructions, real-time feedback, and a structured quest-based learning environment.
]{}

\section{Discussion}
\label{sec:discussion}

This section highlights the insights gained from our research findings and explores potential opportunities for further investigation.


\textbf{Guiding Students with Quests.} Our study found that using OSSDoorway's structured, sequential tasks helped mitigate navigation and task clarity challenges. This aligns with previous research, which shows that step-by-step guidance, such as that provided by quests and points, can reduce barriers for newcomers in OSS environments~\cite{PS04diniz2017using}. Our participants also highlighted the value of quests in effectively tracking their progress and guiding their navigation through the OSS contribution process. Moreover, similar to the findings from Diniz et al.~\cite{PS04diniz2017using}, quests can effectively orient students toward their first contribution, keeping them engaged throughout their journey.


\textbf{Gamification in OSS education.} Applying game elements to OSS contributions can support contribution. 
Even though gamification can also help newcomers overcome barriers~\cite{steinmacher2015social}, especially those related to process and environment, other barriers can still impact student participation, such as toxic culture and exclusionary practices, as highlighted by Trinkenreich et al.~\cite{trinkenreich2021women}. While gamification can enhance engagement, it must complement broader efforts to address systemic issues in OSS communities. Future research should explore how gamification can help mitigate these barriers.

\textbf{Motivations to contribute to OSS projects.} Previous research has explored the motivations behind developers' contributions to OSS~\cite{lakhani2004open, bitzer2007intrinsic, von2012carrots, gerosa2021shifting}. Developers are extrinsically motivated when they seek external rewards, such as career advancement or financial compensation~\cite{frey1997relationship}. These external motivators can become internalized, leading developers to perceive them as self-regulated behaviors driven by factors like reputation, reciprocity, learning, and personal use~\cite{von2012carrots, deci1987support}. Intrinsic motivations, by contrast, arise from acting for its inherent satisfaction, such as ideology, altruism, or fun~\cite{von2012carrots, ryan2000self}. Gerosa et al.~\cite{gerosa2021shifting} found that intrinsic and internalized extrinsic motivations largely drive contributors, while other studies suggest extrinsic rewards, like potential monetary gain, also play a role~\cite{alexander2002working}. Past work has also investigated students' motivation to contribute to OSS~\cite{silva2020google}. Future research should explore how to motivate students to contribute to OSS projects by aligning gamification with these motivational drivers.

\textbf{Students' self-efficacy improvement.} Previous work~\cite{mendez2018open, padala2020gender} has demonstrated that the information architecture of OSS project pages (e.g., project descriptions and issue tracker details) typically appeals to individuals with high self-efficacy, driven by motivations like intellectual stimulation, competition, and learning technology for enjoyment. Our findings indicate that using OSSDoorway significantly enhanced students' self-efficacy in completing tasks related to contributing to OSS projects. These results align with previous studies that demonstrated how redesigning GitHub through a web browser plugin also increased students' self-efficacy~\cite{santos2023designing}. Future research could explore additional strategies to boost students' confidence, such as refining gamified elements or integrating personalized feedback mechanisms to better support learners in the OSS contribution process.

\section{Implications}
\label{sec:implications}

In this section, we outline the implications of our study.
\textbf{\textit{Implications for students.}} For students, especially those with lower self-efficacy, OSSDoorway's structured environment significantly increases self-efficacy in contributing to OSS projects. The results show that quests and real-time feedback help reduce the fear of making mistakes and provide a clear path to contribute to OSS projects. Students can benefit from these features by following guided steps to learn concepts in OSS tools, such as issue tracking and pull requests. Our findings emphasize the need to provide students with an environment that offers support through feedback and structured questions, which can alleviate common frustrations and help students build the skills necessary to contribute effectively.

\textbf{\textit{Implications for educators.}} Incorporating OSS contributions into the computer science curriculum offers valuable real-world experience to students. Still, educators should recognize that some students—particularly those with lower self-efficacy—may struggle more when using platforms like GitHub. Educators could introduce tools like OSSDoorway to provide a more scaffolded learning experience, offering clear, step-by-step tasks and real-time feedback to ease students into OSS contributions. 

\textbf{\textit{Implications for social coding platforms.}} Our findings indicate that structured guidance, such as quests and real-time feedback, benefits students, increasing their self-efficacy. Social coding platforms like GitHub could incorporate similar features to support new users. 

\textbf{\textit{Implications for maintainers of OSS projects.}} 
Projects could benefit from student contributions, especially those who struggle to attract external contributors. Including contextualized quests or guided tutorials could also help lower the entry barriers for students. Additionally, maintainers should consider providing real-time feedback with automated bots, similar to OSSDoorway, to guide new contributors through their first contributions. 

\textbf{\textit{Implications for researchers.}} Our study adds to the current literature on gamification in OSS by exploring its impact on self-efficacy and supporting students in contributing to OSS projects. Future research should further investigate how specific game elements, like quests and feedback systems, affect different students' learning styles and motivations. Additionally, understanding the long-term effects of using gamified tools like OSSDoorway on student retention and performance in real-world OSS projects would be valuable.

\section{Limitations}
\label{sec:threatstovalidity}


Even though we developed OSSDoorway based on the outcomes of each step in our research method and conceptualized this environment using surveys, and interviews, the requirements elicitation process may not fully cover all the nuances of contributing to OSS projects, particularly in capturing various contribution types and strategies. To address this, we evaluated the tasks implemented in OSSDoorway with experts. Nevertheless, while OSSDoorway aims to support students in the initial steps of the contribution process, such as requiring students to make non-code contributions, it does not encompass the full range of contribution methods or delve deeply into technical specifics. Our focus was primarily on the contribution process and the tools rather than the technical details of coding or development. Future studies can explore how OSSDoorway can support non-technical contributors (e.g., writers) in engaging with OSS projects.

While OSS promotes teamwork, gamification can sometimes prioritize competition and personal milestones, potentially overlooking the social and OSS cooperative aspects. In this version of OSSDoorway, we chose not to incorporate competitive elements to ensure the primary focus remained on learning and skill development. However, future studies could explore how gamified approaches might better integrate collaborative and competitive mechanics.

Concerning the generalizability of our study, our results can be partially influenced by the specific sample of participants. OSSDoorway may yield different outcomes if used by students with varying skill levels in OSS tools. Additionally, our study is limited to the use of GitHub. We cannot assume that the benefits observed with OSSDoorway will be as effective in real-world OSS projects as they were identified in the controlled sandbox environment of OSSDoorway. Future work can investigate how students transfer the acquired knowledge to real project settings.

Regarding the reliability of the study's conclusions, we employed nonparametric statistical tests with minimal statistical assumptions and Bonferroni correction for multiple comparisons. For the qualitative analysis, we acknowledge the potential for bias in data interpretation. To mitigate subjectivity, the team engaged in continuous comparative analysis and reached conclusions through a process of negotiated agreement. Each team member has extensive experience in qualitative methods and OSS.

\section{Conclusion}
\label{sec:conclusion}

In this study, we presented OSSDoorway, a gamified environment designed to support students in contributing to OSS projects, and evaluated it with 29 graduate students. Our study demonstrates that OSSDoorway effectively supports students in contributing to OSS projects by significantly improving their self-efficacy and providing a structured, gamified learning environment. The combination of game elements, real-time feedback from the OSS Bot, and the quest-based system helped students navigate complex tasks such as understanding GitHub features, submitting pull requests, and collaborating within a community. Our findings suggest that OSSDoorway can be a valuable tool for supporting students in contributing to OSS projects. Future work will focus on refining and expanding OSSDoorway to include more advanced contributions, such as code review and refactoring. Additionally, we plan to conduct an empirical study to evaluate the OSSDoorway's impact, considering students' different cognitive styles.

\section{Data Availability}

The replication package, with the research instruments and scripts, is available for public access~\cite{replicationpackage}.

\section*{Acknowledgments}

The National Science Foundation (NSF) partially supports this work under grant numbers 2236198, 2247929, 2303042, and 2303612. Katia Felizardo is funded by a research grant from the Brazilian National Council for Scientific and Technological Development (CNPq), Grant 302339/2022--1. We also thank the students for their participation in our study, Connor Ailton, Karissa Smallwood, Jadyn Calhoun, Pedro Oliveira, Kristiana Kirk, and Aaron Santiago for developing OSSDoorway and contributing to the project.


\bibliographystyle{IEEEtran}
\bibliography{IEEEabrv,bibtex.bib}

\begin{thebibliography}{10}
\providecommand{\url}[1]{#1}
\csname url@samestyle\endcsname
\providecommand{\newblock}{\relax}
\providecommand{\bibinfo}[2]{#2}
\providecommand{\BIBentrySTDinterwordspacing}{\spaceskip=0pt\relax}
\providecommand{\BIBentryALTinterwordstretchfactor}{4}
\providecommand{\BIBentryALTinterwordspacing}{\spaceskip=\fontdimen2\font plus
\BIBentryALTinterwordstretchfactor\fontdimen3\font minus \fontdimen4\font\relax}
\providecommand{\BIBforeignlanguage}[2]{{%
\expandafter\ifx\csname l@#1\endcsname\relax
\typeout{** WARNING: IEEEtran.bst: No hyphenation pattern has been}%
\typeout{** loaded for the language `#1'. Using the pattern for}%
\typeout{** the default language instead.}%
\else
\language=\csname l@#1\endcsname
\fi
#2}}
\providecommand{\BIBdecl}{\relax}
\BIBdecl

\bibitem{sedelmaier2012research}
Y.~Sedelmaier and D.~Landes, ``A research agenda for identifying and developing required competencies in software engineering,'' in \emph{2012 15th International Conference on Interactive Collaborative Learning (ICL)}.\hskip 1em plus 0.5em minus 0.4em\relax IEEE, 2012, pp. 1--5.

\bibitem{cantone2014agile}
G.~Cantone, M.~Marchesi \emph{et~al.}, ``Agile processes in software engineering and extreme programming,'' in \emph{Proceedings of 15th international conference, XP}.\hskip 1em plus 0.5em minus 0.4em\relax Springer, 2014.

\bibitem{holmes2018dimensions}
R.~Holmes, M.~Allen, and M.~Craig, ``Dimensions of experientialism for software engineering education,'' in \emph{Proceedings of the 40th International Conference on Software Engineering: Software Engineering Education and Training}, 2018, pp. 31--39.

\bibitem{cawley2014incorporating}
O.~Cawley, S.~Weibelzahl, I.~Richardson, and Y.~Delaney, ``Incorporating a self-directed learning pedagogy in the computing classroom: Problem-based learning as a means to improving software engineering learning outcomes,'' in \emph{Overcoming Challenges in Software Engineering Education: Delivering Non-Technical Knowledge and Skills}.\hskip 1em plus 0.5em minus 0.4em\relax IGI Global, 2014, pp. 348--371.

\bibitem{nascimento2013using}
D.~M. Nascimento, K.~Cox, T.~Almeida, W.~Sampaio, R.~A. Bittencourt, R.~Souza, and C.~Chavez, ``Using open source projects in software engineering education: A systematic mapping study,'' in \emph{2013 IEEE Frontiers in Education Conference (FIE)}.\hskip 1em plus 0.5em minus 0.4em\relax IEEE, 2013, pp. 1837--1843.

\bibitem{pinto2017training}
G.~H.~L. Pinto, F.~Figueira~Filho, I.~Steinmacher, and M.~A. Gerosa, ``Training software engineers using open-source software: the professors' perspective,'' in \emph{CSEE\&T}.\hskip 1em plus 0.5em minus 0.4em\relax IEEE, 2017, pp. 117--121.

\bibitem{silva2020google}
J.~O. Silva, I.~Wiese, D.~M. German, C.~Treude, M.~A. Gerosa, and I.~Steinmacher, ``Google summer of code: Student motivations and contributions,'' \emph{JSS}, vol. 162, p. 110487, 2020.

\bibitem{pinto2019training}
G.~Pinto, C.~Ferreira, C.~Souza, I.~Steinmacher, and P.~Meirelles, ``Training software engineers using open-source software: the students' perspective,'' in \emph{ICSE-SEET 2019}.\hskip 1em plus 0.5em minus 0.4em\relax IEEE, 2019.

\bibitem{braught2018multi}
G.~Braught, J.~Maccormick, J.~Bowring, Q.~Burke, B.~Cutler, D.~Goldschmidt, M.~Krishnamoorthy, W.~Turner, S.~Huss-Lederman, B.~Mackellar \emph{et~al.}, ``A multi-institutional perspective on {H/FOSS} projects in the computing curriculum,'' \emph{ACM Transactions on Computing Education (TOCE)}, vol.~18, no.~2, pp. 1--31, 2018.

\bibitem{morgan2014lessons}
B.~Morgan and C.~Jensen, ``Lessons learned from teaching open source software development,'' in \emph{IFIP International Conference on Open Source Systems}.\hskip 1em plus 0.5em minus 0.4em\relax Springer, 2014.

\bibitem{cai2016reputation}
Y.~Cai and D.~Zhu, ``Reputation in an open source software community: antecedents and impacts,'' \emph{Decision Support Systems}, 2016.

\bibitem{riehle2015open}
D.~Riehle, ``How open source is changing the software developer's career.'' \emph{Computer}, 2015.

\bibitem{Singer2013CSCW}
L.~Singer, F.~Figueira~Filho, B.~Cleary, C.~Treude, M.-A. Storey, and K.~Schneider, ``Mutual assessment in the social programmer ecosystem: An empirical investigation of developer profile aggregators,'' in \emph{Proceedings of the 2013 Conference on Computer Supported Cooperative Work}, ser. CSCW '13.\hskip 1em plus 0.5em minus 0.4em\relax New York, NY, USA: ACM, 2013, pp. 103--116.

\bibitem{Marlow2013CSCW}
J.~Marlow, L.~Dabbish, and J.~Herbsleb, ``Impression formation in online peer production: Activity traces and personal profiles in {GitHub},'' in \emph{Proceedings of the 2013 Conference on Computer Supported Cooperative Work}, ser. CSCW '13.\hskip 1em plus 0.5em minus 0.4em\relax New York, NY, USA: ACM, 2013, pp. 117--128.

\bibitem{parra2016making}
E.~Parra, S.~Haiduc, and R.~James, ``Making a difference: an overview of humanitarian free open source systems,'' in \emph{ICSE 2016-Companion}.\hskip 1em plus 0.5em minus 0.4em\relax {ACM}, 2016.

\bibitem{greene2016cvexplorer}
G.~J. Greene and B.~Fischer, ``Cvexplorer: identifying candidate developers by mining and exploring their open source contributions,'' in \emph{ASE 2016}.\hskip 1em plus 0.5em minus 0.4em\relax {ACM}, 2016.

\bibitem{steinmacher2015social}
I.~Steinmacher, T.~Conte, M.~Gerosa, and D.~Redmiles, ``Social barriers faced by newcomers placing their first contribution in open source software projects,'' in \emph{ACM CSCW 2015}, 2015.

\bibitem{mendez2018open}
C.~Mendez, H.~S. Padala, Z.~Steine-Hanson, C.~Hilderbrand, A.~Horvath, C.~Hill, L.~Simpson, N.~Patil, A.~Sarma, and M.~Burnett, ``Open source barriers to entry, revisited: a sociotechnical perspective,'' in \emph{ICSE 2018}, 2018.

\bibitem{padala2020gender}
S.~H. Padala, C.~J. Mendez, L.~F. Dias, I.~Steinmacher, Z.~S. Hanson, C.~Hilderbrand, A.~Horvath, C.~Hill, L.~D. Simpson, M.~Burnett \emph{et~al.}, ``How gender-biased tools shape newcomer experiences in {OSS} projects,'' \emph{IEEE TSE}, 2020.

\bibitem{trinkenreich2021women}
B.~Trinkenreich, I.~Wiese, A.~Sarma, M.~Gerosa, and I.~Steinmacher, ``Women's participation in open source software: a survey of the literature,'' \emph{ACM TOSEM}, 2022.

\bibitem{steinmacher2016overcoming}
I.~Steinmacher, T.~Conte, C.~Treude, and M.~Gerosa, ``Overcoming open source project entry barriers with a portal for newcomers,'' in \emph{ICSE 2016}, 2016.

\bibitem{santos2022hits}
I.~Santos, I.~Wiese, I.~Steinmacher, A.~Sarma, and M.~A. Gerosa, ``Hits and misses: Newcomers' ability to identify skills needed for {OSS} tasks,'' in \emph{IEEE SANER}.\hskip 1em plus 0.5em minus 0.4em\relax IEEE, 2022, pp. 174--183.

\bibitem{dichev2017gamifying}
C.~Dichev and D.~Dicheva, ``Gamifying education: what is known, what is believed and what remains uncertain: a critical review,'' \emph{International journal of educational technology in higher education}, vol.~14, pp. 1--36, 2017.

\bibitem{pedreira2015gamification}
O.~Pedreira, F.~Garc{\'\i}a, N.~Brisaboa, and M.~Piattini, ``Gamification in software engineering--a systematic mapping,'' \emph{Information and Software Technology (IST)}, vol.~57, pp. 157--168, 2015.

\bibitem{dicheva2015gamification}
D.~Dicheva, C.~Dichev, G.~Agre, and G.~Angelova, ``Gamification in education: A systematic mapping study,'' \emph{ET\&S}, vol.~18, no.~3, pp. 75--88, 2015.

\bibitem{bartel2016gamifying}
A.~Bartel and G.~Hagel, ``Gamifying the learning of design patterns in software engineering education,'' in \emph{EDUCON}.\hskip 1em plus 0.5em minus 0.4em\relax IEEE, 2016, pp. 74--79.

\bibitem{PS04diniz2017using}
G.~C. Diniz, M.~A.~G. Silva, M.~A. Gerosa, and I.~Steinmacher, ``Using gamification to orient and motivate students to contribute to {OSS} projects,'' in \emph{IEEE/ACM 10th CHASE}.\hskip 1em plus 0.5em minus 0.4em\relax IEEE, 2017, pp. 36--42.

\bibitem{Belland2014}
B.~R. Belland, \emph{Scaffolding: Definition, Current Debates, and Future Directions}.\hskip 1em plus 0.5em minus 0.4em\relax New York, NY: Springer New York, 2014, pp. 505--518.

\bibitem{deterding2011gamification}
S.~Deterding, R.~Khaled, L.~E. Nacke, D.~Dixon \emph{et~al.}, ``Gamification: Toward a definition,'' in \emph{CHI}, vol.~12.\hskip 1em plus 0.5em minus 0.4em\relax Vancouver, 2011, pp. 1--79.

\bibitem{sailer2020gamification}
M.~Sailer and L.~Homner, ``The gamification of learning: A meta-analysis,'' \emph{Educational psychology review}, vol.~32, no.~1, pp. 77--112, 2020.

\bibitem{marin2018empirical}
B.~Mar{\'\i}n, J.~Frez, J.~Cruz-Lemus, and M.~Genero, ``An empirical investigation on the benefits of gamification in programming courses,'' \emph{ACM Transactions on Computing Education (TOCE)}, vol.~19, no.~1, pp. 1--22, 2018.

\bibitem{ayub2019gamification}
M.~Ayub, H.~Toba, M.~C. Wijanto, S.~Yong, and B.~Wijaya, ``Gamification for blended learning in higher education,'' \emph{World Transactions on Engineering and Technology Education}, vol.~17, no.~1, pp. 76--81, 2019.

\bibitem{toda2019analysing}
A.~M. Toda, A.~C. Klock, W.~Oliveira, P.~T. Palomino, L.~Rodrigues, L.~Shi, I.~Bittencourt, I.~Gasparini, S.~Isotani, and A.~I. Cristea, ``Analysing gamification elements in educational environments using an existing gamification taxonomy,'' \emph{Smart Learning Environments}, vol.~6, no.~1, pp. 1--14, 2019.

\bibitem{prause2012field}
C.~R. Prause, J.~Nonnen, and M.~Vinkovits, ``A field experiment on gamification of code quality in agile development.'' in \emph{PPIG}, 2012, p.~17.

\bibitem{rojas2016teaching}
J.~M. Rojas and G.~Fraser, ``Teaching mutation testing using gamification,'' in \emph{ECSEE}, 2016.

\bibitem{garaccione2022gerry}
G.~Garaccione, T.~Fulcini, and M.~Torchiano, ``Gerry: a gamified browser tool for {GUI} testing,'' in \emph{Gamify'22}, 2022, pp. 2--9.

\bibitem{ozturk2022gamification}
S.~Ozturk, ``Gamification of exploratory testing process,'' in \emph{Gamify'22}, 2022, pp. 14--17.

\bibitem{annunziata2024serge}
G.~Annunziata, S.~Lambiase, F.~Palomba, and F.~Ferrucci, ``Serge--serious game for the education of risk management in software project management,'' in \emph{Proceedings of the 46th International Conference on Software Engineering: Software Engineering Education and Training}, 2024, pp. 264--273.

\bibitem{navarro2007comprehensive}
E.~O. Navarro and A.~Van Der~Hoek, ``Comprehensive evaluation of an educational software engineering simulation environment,'' in \emph{20th Conference on Software Engineering Education \& Training (CSEET'07)}.\hskip 1em plus 0.5em minus 0.4em\relax IEEE, 2007, pp. 195--202.

\bibitem{garaccione2024gamification}
G.~Garaccione, R.~Coppola, L.~Ardito, and M.~Torchiano, ``Gamification of business process modeling education: an experimental analysis,'' \emph{Software and Systems Modeling}, pp. 1--26, 2024.

\bibitem{su2016effects}
C.-H. Su, ``The effects of students' motivation, cognitive load and learning anxiety in gamification software engineering education: a structural equation modeling study,'' \emph{Multimedia Tools and Applications}, vol.~75, pp. 10\,013--10\,036, 2016.

\bibitem{sheth2012increasing}
S.~K. Sheth, J.~S. Bell, and G.~E. Kaiser, ``Increasing student engagement in software engineering with gamification,'' in \emph{4th SSE}, 2012.

\bibitem{zhao152023motivating}
S.~Zhao15, X.~Xia25, B.~Fitzgerald, X.~Li, V.~Lenarduzzi, D.~Taibi, R.~Wang, W.~Wang, and C.~Tian, ``Motivating open source collaborations through social network evaluation: A gamification practice from alibaba,'' in \emph{IEEE/ACM 46th ICSE-SEIP}, 2024.

\bibitem{santoslr}
I.~Santos, K.~R. Felizardo, I.~Steinmacher, and M.~A. Gerosa, ``Software solutions for newcomers’ onboarding in software projects: A systematic literature review,'' \emph{Information and Software Technology}, vol. 177, p. 107568, 2025.

\bibitem{PS17toscani2018gamification}
C.~Toscani, D.~Gery, I.~Steinmacher, and S.~Marczak, ``A gamification proposal to support the onboarding of newcomers in the flosscoach portal,'' in \emph{17th IHC}, 2018, pp. 1--10.

\bibitem{santosgamelements}
I.~Santos, K.~R. Felizardo, M.~A. Gerosa, and I.~Steinmacher, ``Game elements to engage students learning the open source software contribution process,'' in \emph{2024 IEEE Symposium on Visual Languages and Human-Centric Computing (VL/HCC)}, 2024, pp. 59--70.

\bibitem{rosner2006wilcoxon}
B.~Rosner, R.~J. Glynn, and M.-L.~T. Lee, ``The wilcoxon signed rank test for paired comparisons of clustered data,'' \emph{Biometrics}, 2006.

\bibitem{strauss1998basics}
A.~Strauss and J.~Corbin, \emph{Basics of qualitative research techniques}.\hskip 1em plus 0.5em minus 0.4em\relax Sage Publications, 1998.

\bibitem{eccles2017think}
D.~W. Eccles and G.~Arsal, ``The think aloud method: what is it and how do i use it?'' \emph{Qualitative Research in Sport, Exercise and Health}, vol.~9, no.~4, pp. 514--531, 2017.

\bibitem{trinkenreich2020hidden}
B.~Trinkenreich, M.~Guizani, I.~Wiese, A.~Sarma, and I.~Steinmacher, ``Hidden figures: Roles and pathways of successful oss contributors,'' \emph{ACM on Human-Computer Interaction}, vol.~4, no. CSCW, pp. 1--22, 2020.

\bibitem{nodejs}
\BIBentryALTinterwordspacing
N.~Foundation, ``Node{JS},'' 2024, accessed: 2024-09-20. [Online]. Available: \url{https://nodejs.org/}
\BIBentrySTDinterwordspacing

\bibitem{probot}
\BIBentryALTinterwordspacing
P.~Team, ``Probot framework,'' 2024, accessed: 2024-09-20. [Online]. Available: \url{https://probot.github.io/}
\BIBentrySTDinterwordspacing

\bibitem{burnett2018gendermag}
M.~Burnett, S.~Stumpf, L.~Beckwith, and A.~Peters, ``The {GenderMag} kit: how to use the {GenderMag} method to find inclusiveness issues through a gender lens,'' 2018.

\bibitem{replicationpackage}
\BIBentryALTinterwordspacing
Figshare, ``Replication package,'' 2024, accessed: 2024-09-09. [Online]. Available: \url{https://figshare.com/s/0b93fef5c7ef015e2ddf}
\BIBentrySTDinterwordspacing

\bibitem{bandura2014social}
A.~Bandura, ``Social cognitive theory of moral thought and action,'' in \emph{Handbook of Moral Behavior and Development}.\hskip 1em plus 0.5em minus 0.4em\relax Psychology press, 2014.

\bibitem{macfarland2016mann}
T.~W. MacFarland, J.~M. Yates, T.~W. MacFarland, and J.~M. Yates, ``Mann--whitney u test,'' \emph{Introduction to nonparametric statistics for the biological sciences using R}, pp. 103--132, 2016.

\bibitem{lakhani2004open}
K.~R. Lakhani and E.~Von~Hippel, \emph{How open source software works:“free” user-to-user assistance}.\hskip 1em plus 0.5em minus 0.4em\relax Springer, 2004.

\bibitem{bitzer2007intrinsic}
J.~Bitzer, W.~Schrettl, and P.~J. Schr{\"o}der, ``Intrinsic motivation in open source software development,'' \emph{Journal of comparative economics}, vol.~35, no.~1, pp. 160--169, 2007.

\bibitem{von2012carrots}
G.~Von~Krogh, S.~Haefliger, S.~Spaeth, and M.~W. Wallin, ``Carrots and rainbows: motivation and social practice in open source software development,'' \emph{MIS Q}, 2012.

\bibitem{gerosa2021shifting}
M.~Gerosa, I.~Wiese, B.~Trinkenreich, G.~Link, G.~Robles, C.~Treude, I.~Steinmacher, and A.~Sarma, ``The shifting sands of motivation: Revisiting what drives contributors in open source,'' in \emph{ICSE 2021}.\hskip 1em plus 0.5em minus 0.4em\relax {IEEE}, 2021.

\bibitem{frey1997relationship}
B.~S. Frey \emph{et~al.}, ``On the relationship between intrinsic and extrinsic work motivation,'' \emph{International journal of industrial organization}, vol.~15, no.~4, pp. 427--439, 1997.

\bibitem{deci1987support}
E.~L. Deci and R.~M. Ryan, ``The support of autonomy and the control of behavior.'' \emph{Journal of personality and social psychology}, vol.~53, no.~6, p. 1024, 1987.

\bibitem{ryan2000self}
R.~M. Ryan and E.~L. Deci, ``Self-determination theory and the facilitation of intrinsic motivation, social development, and well-being.'' \emph{American psychologist}, vol.~55, no.~1, p.~68, 2000.

\bibitem{alexander2002working}
S.~O. Alexander~Hars, ``Working for free? motivations for participating in open-source projects,'' \emph{International journal of electronic commerce}, vol.~6, no.~3, pp. 25--39, 2002.

\bibitem{santos2023designing}
I.~Santos, J.~F. Pimentel, I.~Wiese, I.~Steinmacher, A.~Sarma, and M.~A. Gerosa, ``Designing for cognitive diversity: Improving the {GitHub} experience for newcomers,'' in \emph{2023 IEEE/ACM 45th ICSE-SEIS}.\hskip 1em plus 0.5em minus 0.4em\relax IEEE, 2023, pp. 1--12.

\end{thebibliography}

\end{document}